\documentclass[aps,twocolumn,secnumarabic,amsmath,amssymb]{revtex4-2}
\usepackage{graphicx} 
\usepackage[letterpaper,top=0.75in,left=0.75in,bottom=1in,right=0.75in]{geometry}
\usepackage[utf8]{inputenc} 
\usepackage{gensymb}
\usepackage[colorlinks=true]{hyperref}
\usepackage{tabularx}

\begin{document}

\title{\LARGE AI in Software Engineering: A Survey on Project Management Applications}

\author{Talia Crawford, Scott Duong, Richard Fueston, Ayorinde Lawani, Samuel Owoade, Abel Uzoka, Reza M. Parizi}

\affiliation{Department of Software Engineering and Game Development, Kennesaw State University, Marietta, GA, USA\\
\{tcrawf28, sduong3, rfueston , alawani, sowoade, auzoka\}@students.kennesaw.edu, rparizi1@kennesaw.edu}

\author{Abbas Yazdinejad}
\affiliation{Cyber Science Lab, School of Computer Science, University of Guelph, Ontario, Canada\\ayazdine@uoguelph.ca}


\maketitle

\textbf{\textit{Abstract}  --  Artificial Intelligence (AI) refers to the intelligence demonstrated by machines, and within the realm of AI, Machine Learning (ML) stands as a notable subset. ML employs algorithms that undergo training on data sets, enabling them to carry out specific tasks autonomously. Notably, AI holds immense potential in the field of software engineering, particularly in project management and planning. In this literature survey, we explore the use of AI in Software Engineering and summarize previous works in this area. We first review eleven different publications related to this subject, then compare the surveyed works. We then comment on the possible challenges present in the utilization of AI in software engineering and suggest possible further research avenues and the ways in which AI could evolve with software engineering in the future. }


\section{Introduction}
Artificial Intelligence (AI) (including Machine Learning (ML)) is an emerging part of the software engineering landscape. Much of the success of software projects depends heavily on project management and preparation. A team's success can be influenced by whether the project was adequately planned. Project requirements, cost estimation, and risk mitigation are some of these crucial aspects. Mistakes in this type of decision can result in increased costs or project failure. Therefore, many software engineers are now looking for ways in which AI could be applied to reduce the error involved in project development decisions \cite{a1}. 

Many are hopeful that AI can be utilized in the near future to manage potential risks and organize projects more effectively. If AI is successfully implemented in software project development decisions, there could be a decrease in project failures and errors during development and an increase in project efficiency and quality. The way in which AI/ML is intended to be used in current software engineering is shown in the following Figure \ref{fig:currentai}. 

\begin{figure}[h]
    \centering
    \includegraphics[width=80mm]{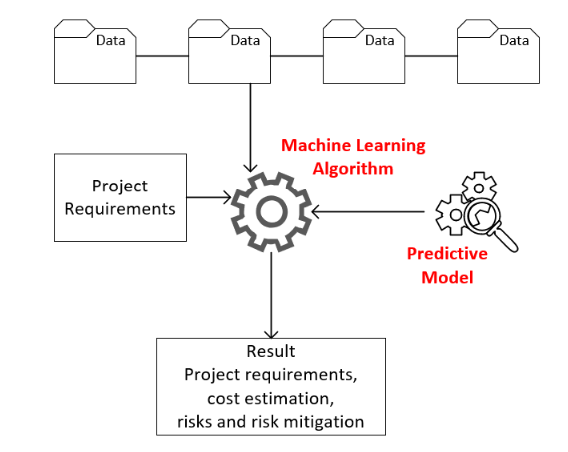}
    \caption{This figure shows how current AI and ML work with software engineering.}
    \label{fig:currentai}
\end{figure}

\section{Surveyed Works} \label{survey}
There are many previously published works related to Artificial Intelligence in Software Engineering. Here we will review eleven such publications that discuss and are related to the intersection of software engineering and artificial intelligence. Each of these publications is unique, with the oldest being from 1992 and the most recent being from 2022. We decided to include the sources from 1992 as a way to see the larger picture involved in implementing AI in software engineering. While these two sources do not reflect the current attitude and ideas surrounding this intersection of AI and software engineering, they provide an important perspective so that we can see how the landscape has changed and how it has remained the same over the last 30 years of development. 

The other nine sources that were surveyed were published between 2018 and 2022, which we believe provides a range that covers the current information on this subject as it is rapidly changing. 
In the paper ``Towards Effective AI-powered Agile Project Management"\cite{dam}, Dam \textit{et al.} propose a framework for how AI technology can be used to support agile project management. Their proposed framework assists in certain aspects of agile project management, such as ``automating repetitive high-volume tasks to enable project analytics for estimation and risk prediction" \cite{dam}. In ``Artificial Neural Network Architecture and Orthogonal Arrays in Estimation of Software Projects Efforts"\cite{rankovic}, Ranković \textit{et al.} propose a model of an artificial neural network to minimize error in software project development estimation. They combined both traditional mathematical models as well as machine learning algorithms in order to improve the accuracy of the assessment of effort and cost of developing a software system \cite{rankovic}. 
In Khanna \textit{et al.}'s ``Artificial Intelligence based Risk Management Framework for Distributed Agile Software Development"\cite{khanna}, the authors propose a system to mitigate risks in Distributed Agile Software Development. They first address the differences between distributed development and agile development and then propose a way to mitigate risks for the combined development system. The authors state that ``manual risk management is ineffective as it is purely dependent on the experience and analysis of the person conducting it," and therefore there is a need for automated risk assessment and prioritization \cite{khanna}.

In ``A Tool For Software Requirement Allocation Using Artificial Intelligence Planning"\cite{pereira}, Pereira \textit{et al.} introduce a plan for an AI task allocation tool in order to allocate software requirements into versions. Their proposed tool reviews the requirements considering the development time, priority levels, and dependency relationships in order to allocate each requirement. They identified the main limitation of this proposed model as the ``subjectivity of identifying values used as requirement priority, size, and dependencies."\cite{pereira}. In the article ``Machine Learning for Software Engineering: Models, Methods, and Applications"\cite{meinke}, Meinke and Bennaceur look at the application of ML in software engineering. They state that while machine learning is a mature discipline, perspectives of ML in software engineering are much less common. Machine learning has the potential to become very useful to the software engineering discipline to coincide with the current upwards trend of agile development, as ML has the ability to meet the needs of these changing approaches. \cite{meinke}. 
One of the earliest publications on AI use in software engineering is "The Relationships of AI to software engineering," a talk by Partridge from 1992. Partridge discusses having AI provide support for the software development process and decision-making \cite{partridge}. This is in line with current ideas of how AI would best be used in software engineering. 

Another talk from the same conference in 1992 provides us with a second of the earliest publications on AI in software engineering. ``Application of AI and model building techniques to software engineering,"\cite{bennett} is a publication by Bennett \textit{et al.} that details work in using AI to support modeling and simulation of dynamical systems, as a part of which they considered how ``the techniques can be applied to model-based software engineering methodologies for specification and design of software"\cite{bennett}. This source and the previous source provide an important place of reference for us to understand that this idea of using AI for software engineering and project management is not unique to the current trend of agile development, and software engineers have been seeing the need for support in decision making for many years. 
In the article ``Software Engineering for Machine-Learning Applications: The Road Ahead"\cite{khomh}, Khomh \textit{et al.} discuss the challenges of using AI and ML in software engineering. The main challenge is that it can be difficult to effectively develop, test, and evolve a system since there may not be complete specifications for some of the most important behaviors.  The author Lu \textit{et al.} discusses the ability of AI to make decisions in an ethical way in ``Software Engineering for Responsible AI: An empirical study and operationalized patterns"\cite{lu}. This paper discusses the different ethical principles involved in using AI models. Since AI is complex, it makes risk assessment of its adoption very difficult, and there are many trade-offs between the different principles. Dynamic and adaptive ethics assessment needs to be conducted for each context in which AI is proposed to be used. AI systems should be integrated into requirement-driven and outcome-driven development, as they continuously learn based on new data. These requirements need to be specified, verifiable, and measurable to ensure accurate requirements validation.

In the paper ``Analysis of Software Engineering for Agile Machine Learning Projects"\cite{singla}, Singla \textit{et al.} look at the issues in tracking data from machine learning projects in the scrum. They analyze the types of issues found in the collected data and propose some ways in which Agile machine learning projects can improve their logging and execution\cite{singla}.
The paper, ``The Future of Software Engineering by 2050s: Will AI Replace Software Engineers?" seeks to establish the state of the software engineering role by 2050 with regard to inevitable changes necessitated by AI. Basing the work of a survey of software engineers in the field, the author argues that the engineer's role might be different, especially in the design, coding, and testing phases of the project. The work concludes that while some software engineering roles could be replaced by AI, new roles are going to be created - in AI systems moderation and development. \cite{zohair}

\begin{table*}[!ht]
    \caption{ A comparison of ten surveyed sources as previously reviewed in Section \ref{survey}. }
    \label{tab:table1}
    \centering
   \begin{tabular}{|p{.7cm}|p{3cm}|p{1cm}|p{4cm}|p{4.5cm}|}
        \hline
        \textbf{Cit. No.} & \textbf{Authors} & \textbf{Year} & \textbf{Type of Publication}  & \textbf{Limitations} \\
        \hline
        \cite{dam} & H. K. Dam, T. Tran, J. Grundy, A. Ghose and Y. Kamei & 2019 & Proposal of an AI System for Agile PM  & This proposal plans to develop a prototype and test it on 150 open source projects. This can be limiting in that the projects are restricted to only open source projects that are available to the authors.  \\
        \hline
        \cite{rankovic} & N. Ranković, D. Ranković, M. Ivanović and L. Lazić & 2021 & Proposal for Artificial Neural Network use in Software Engineering risk estimation  & This publication uses a limited sample size, but takes many steps in their approach to minimize limitations. \\
        \hline
        \cite{khanna} & E. Khanna, R. Popli and N. Chauhan & 2021 & Proposed framework for AI use in Distributed Agile software development methodology  & This includes a proposed framework based on a survey of other literature, where limitations can arise due to the limited number of other papers they were able to survey when creating their own framework proposal. \\
        \hline
        \cite{pereira} & F. C. Pereira, G. B. Neto, L. F. De Lima, F. Silva and L. M. Peres & 2022 & Proposal of an AI tool for software requirement allocation  & This is a validation study and external factors can greatly influence the results. The main limitation is the ``subjectivity of identifying values used as requirement priority, requirement size, and dependencies."\cite{pereira} \\
        \hline
        \cite{meinke} & K. Meinke and A. Bennaceur & 2018 & Proposed technical briefing of Machine Learning in software engineering development  & Lack of successful examples of ML working in a SE project. \\
        \hline
        \cite{partridge} & D. Partridge & 1992 & Extended abstract of a talk given on the interactions of AI to software engineering  & This source is from 1992 so it doesn't include current information. \\
        \hline
        \cite{bennett} & S. Bennett, D. A. Linkens, E. B. Tanyi and A. Scott & 1992 & Extended abstract of a talk given on the use of AI techniques to support software engineering  & This source is from 1992 so it doesn't include current information. \\
        \hline
        \cite{khomh} & F. Khomh, B. Adams, J. Cheng, M. Fokaefs and G. Antoniol & 2018 & Account of possible challenges or solutions of merging ML and Software Engineering as brought up during a conference  & Each of the mentioned challenges or solutions were brought up at a small conference. \\
        \hline
        \cite{lu} & Q. Lu, L. Zhu, X. Xu, J. Whittle, D. Douglas and C. Sanderson & 2022 & Results of interviews conducted with AI/ML scientists and engineers pertaining to ethical issues in AI/ML projects  & The sample size of the interviews conducted were restricted to only responders in Australia. \\
        \hline
        \cite{singla} & K. Singla, J. Bose, and C. Naik & 2018 & Results of a study conducted within a company on using agile methodology with machine learning projects  & The size of this study was small and completely within one company. \\
        \hline
        \cite{zohair} & L. M. Abu Zohair & 2018 & Results of a study conducted using an online survey about the future of AI given to professionals in the field  & The surveyed professionals were limited to those who responded to the online survey within a short time frame. Further views could have been collected to provide more valuable data. \cite{zohair} \\
        \hline
    \end{tabular}
\end{table*}

\section{Comparison of Works} \label{comparison}
In order to best represent the similarities and differences between each of the previously discussed papers, we have included the following Table \ref{tab:table1}. 

This table includes a comparison of the year each source was published, the type of publication, and any limitations mentioned. The type of publication gives information surrounding the publication itself, such as whether it is a proposal for a tool, a conference talk, or a study that was conducted. The limitations included in the following table are either limitations identified in the paper or limitations surrounding the paper's pertinence and usability.

\section{Challenges} \label{challenges}
While most of these sources agree that AI support would be immensely useful in software engineering project management, they also bring up some of the possible challenges in trying to implement AI usage in teams. One of these challenges is the accuracy of systems built using AI and ML models. It has been pointed out that while programs may be able to imitate human behavior, practical products cannot use the same methods that humans use, as they cannot guarantee accuracy \cite{ khomh}. 
Another challenge is the ethics involved in implementing AI models. Currently, most companies utilize a single risk assessment of an AI model and then move on, which is not usually sufficient for a high-level and continuously learning AI system \cite{lu}. There can also be difficulties in testing ML and AI systems. These systems might need a different approach to testing than a regular system since the AI might be incorrect even if the algorithms are correctly implemented. AI systems are only able to learn from available data, so if enough data isn't present it can become very difficult to correctly train and implement an AI system.

One of the greatest challenges to the industry for AI is its ability to behave for the AI \cite{lu}. There have been many chat AI experiments that have gone poorly and had to be taken down. This can be due to the data present for the AI changing the way it works as stated above. This can lead to a lot of issues and catastrophic public relations issues for companies leading to a fear of adoption, as can be seen with companies’ hesitance in ingratiating with ChatGPT. No one in the industry wants to be known for an AI bot that goes crazy, so gaining the trust of the industry is difficult \cite{lu}. As for adoption, there is a challenge in the change in the behavior of people using AI. Once users start trusting AI; they can become accustomed to relying on AI and their behavior in a software engineering setting can change, which could be for the worse as it could also change what the AI learns \cite{khomh,a2}. Challenges for development practices and agile practices can also be an issue that can be affected by AI. For agile scrum practice, it can be extremely important to have team context, and the data given to the AI might not be encompassing enough to count for all aspects, including some of the human work and interaction elements. Agile is already challenging for Trained Scrum masters to ensure that the teams are performing well AI might make bad choices due to lack of data sets and cause sprints to fail. This could be seen in story sizing or grooming task automatically \cite{dam}. As for development tasks, the future of development is most likely very AI driven in its coding tools. There are already a lot of suggested code completions for current IDEs. However, this can be very challenging as developers lose the ability to code with assistance and end up relying on existing code data without the ability to create new ways to solve problems without existing code examples. \cite{zohair}

\section{ Emerging New Technologies in Software Engineering Project Management}
As we explore the confluence of AI and Software Engineering Project Management, it's crucial to consider the transformative impact of other emerging technologies. Blockchain in Software Engineering Project Management, IoT-Enhanced Software Engineering Project Management, and AI-Driven Cybersecurity in Software Development each offer unique solutions and approaches that further strengthen AI's role within this realm. These technologies individually address a myriad of challenges prevalent in project management, simultaneously paving the way for innovative solutions. In this section, we examine how these technologies intersect with AI within software engineering project management, shedding light on their current applications and potential to shape future practices.
\subsection{Blockchain in Software Engineering Project Management}
At its core, blockchain technology provides a decentralized and transparent mechanism for recording transactions. This inherent characteristic of immutability and transparency can be leveraged in software engineering project management to enhance several aspects, particularly concerning trust, collaboration, and traceability \cite{a3}. Project management often involves various stakeholders, including developers, testers, project managers, and clients, among others. Each stakeholder contributes to the project's progress in a unique way, resulting in various activities and transactions. By leveraging blockchain technology, these transactions can be recorded in an immutable and transparent manner. This feature not only enhances trust among stakeholders but also improves auditability \cite{a6}. For instance, task assignments, progress updates, and milestone achievements can be recorded on a blockchain. With each block representing a particular activity, stakeholders can verify the project's state at any given time, enhancing transparency. Blockchain also significantly reduces disputes since every action is traceably documented and cannot be altered retroactively \cite{a4,a5}.

Moreover, blockchain technology can foster efficient and secure collaboration in distributed teams. Smart contracts, a key component of blockchain technology, can automate project management aspects, such as milestone-based payments or resource allocation, based on predefined rules. This automation reduces administrative overhead and allows stakeholders to focus on their primary responsibilities, thereby improving productivity. However, it's important to note that the integration of blockchain technology into software engineering project management is not without challenges. The immutability of blockchain means that any mistake recorded on the blockchain is permanent, which calls for a robust error prevention and mitigation mechanism. Furthermore, privacy concerns, computational overhead, and the complexity of implementing blockchain should also be taken into account while considering this technology for project management \cite{rf2,rf,rf3}.

\subsection{IoT-Enhanced Software Engineering Project Management}

The Internet of Things (IoT), with its capacity to connect billions of devices and facilitate real-time data exchange, can dramatically impact software engineering project management \cite{b1}. IoT can augment AI's capability by providing a wealth of real-time data that can be analyzed and utilized to make better project management decisions. For example, an IoT-enabled workspace could monitor various project-related parameters, such as the developer's working hours, meeting durations, code check-in frequency, and much more. This data, when combined with AI and ML algorithms, can provide insights into project progress, productivity trends, and potential bottlenecks  \cite{b2,b4}. It could also help in predictive tasks such as estimating project completion times or identifying potential risks, thereby enabling proactive project management.
IoT can also enhance collaboration and communication within a project team. Team members can communicate effectively, irrespective of their geographical location, through connected devices and collaborative tools. This is especially crucial in today's age of remote working and distributed teams. Despite the immense potential of IoT in project management, there are certain challenges to be addressed. Data security and privacy are a primary concern, given the sensitive nature of project data. Implementing the appropriate safeguards to protect this data is a must. Additionally, managing the vast amount of data generated by IoT devices requires significant computational resources and efficient data management strategies\cite{b3,b5}.

\subsection{AI-Driven Cybersecurity in Software Development}

AI has been a game-changer in the cybersecurity landscape, and its application in software development is no exception. In the context of software engineering project management, AI-driven cybersecurity can play a pivotal role in ensuring secure development practices and safeguarding project artifacts \cite{c1}. AI and ML algorithms can help identify potential security vulnerabilities in the software being developed. Through the analysis of code patterns, AI can detect anomalies, flag insecure coding practices, and even suggest remediation measures. This proactive approach to security can significantly reduce the chances of security breaches and the costs associated with them. Furthermore, AI can also enhance access control and data protection in project management tools \cite{c3,c2}. By learning the typical access patterns, AI can detect and prevent unauthorized access, thereby protecting sensitive project data. Similarly, AI-driven cybersecurity measures can safeguard sensitive project communications, ensuring that confidential project information is securely transmitted and stored.

In addition to securing the project artifacts, AI can play an essential role in the area of risk management, a key aspect of project management. AI can help identify potential cybersecurity risks at an early stage of the project, allowing project managers to take necessary mitigation measures proactively. While the application of AI in cybersecurity offers numerous benefits, it's not devoid of challenges. A primary concern is the false positives that may result from AI-based threat detection, leading to unnecessary remediation activities. Furthermore, maintaining the privacy of project data while using it to train AI models can also be a challenge \cite{c4}. These emerging technologies, i.e., Blockchain, IoT, and AI-Driven Cybersecurity, have the potential to revolutionize software engineering project management. They can enhance various aspects of project management, including collaboration, transparency, risk management, and security. However, the effective integration of these technologies requires careful consideration of their associated challenges and the development of appropriate strategies to address them. As technology continues to advance, it will be interesting to see how these trends evolve and further shape the future of software engineering project management.

\section{Future Road Map} \label{future}
While there are challenges to using AI in software engineering, there is also possible future work that could mitigate these problems. Firstly, many of the reviewed papers were proposals for systems using AI or ML in software engineering. In order to overcome potential problems, we first need to work on getting basic implementations of AI trained to solve the problems engineers face in project management. Only once we have this baseline will we be able to truly address the issues that arise, logistic, ethical, or otherwise. 

A lot of this will be handled by new and easy-to-use abstraction layers making the future of coding easy for anyone. The key is for developers to cope with the AI changes to prevent being replaced. Development jobs might change into more administrative-style jobs, as can currently be seen in development operations. \cite{zohair}

The majority of the papers surveyed agree that AI would change the nature of a software development project. Data analysis is posited as a major task in which AI/ML would be a factor. Inevitably, AI/ML shall evolve to assist the project manager with more technical tasks - establishing these tasks and training the models are some of the work that we expect to see going forward. Specific areas of future AI/ML applications include project planning, reporting, and monitoring.

\begin{figure}[ht]
    \centering
    \includegraphics[width=80mm]{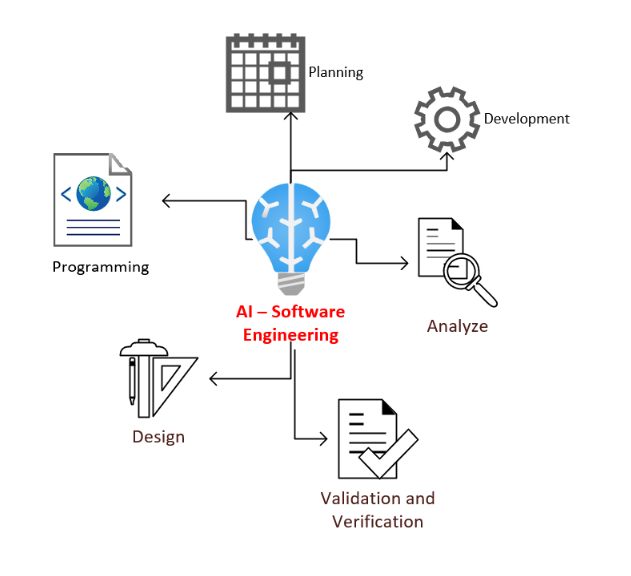}
    \caption{This figure shows how current AI and ML work with software engineering.}
    \label{fig:futureai}
\end{figure}

There is potential for AI to be integrated into more than just the project management area but throughout project development. In Figure \ref{fig:futureai}, we show some possible future uses of AI in software engineering. 

\section{Conclusions} \label{conclusions}
In the field of Software Engineering, Artificial Intelligence, and Machine Learning are extremely rapidly developing technologies. AI and ML have the potential to become invaluable to software engineering, not just in the active parts of software development but also on the project management side. 

In this paper, we have reviewed eleven different sources that all pertain to AI in software engineering. Throughout these sources, we have found commonalities in how AI could be utilized in project management. Some of the main causes of failure in software development are errors made in the project planning phase and mismanagement of projects. If AI were able to step in and reduce some of the uncertainty in this initial decision-making and planning, then software development projects could have higher rates of success. Some specific areas that could greatly improve would be risk assessment and management, cost estimation, and project requirements.

Of our eleven surveyed papers, we chose to review two that were from 1992. We chose to do this so that we could gain a wider perspective on how AI has been developing in the field of software engineering over the past thirty years. These sources allowed us to understand that even when the field of AI was not developed enough to have been a viable option to turn to, and before Agile was the popular framework it is today, software engineers were identifying the same problems and potential AI solutions that we have today.

Eight of our sources were of more current information. They each discussed either possible proposals for AI systems that would improve software engineering project management or results from studies relating to how AI and software engineering are intersecting now. Many sources identified common potential limitations, as well as potential solutions to current project management issues. 

Our last source, Zohair \cite{zohair}, was a survey of professionals in the software field on the future of AI within the next thirty years. This provides a good contrast to how software engineers were viewing AI thirty years ago by showing where software engineers expect AI to be in 2050. Given the inevitability of and how rapidly progressing the field of AI is, it is important to note that not only can AI evolve to provide support for the areas in which project failures are most likely to stem but also in the areas which may not necessarily need AI intervention to prevent failure. While there is the potential for Artificial Intelligence and Machine Learning to take over some aspects of software engineers' jobs, its constantly growing and evolving nature will open up new avenues for software developers to pursue. 

In conclusion, integrating Artificial Intelligence and Machine Learning into the software engineering field can potentially solve many problems that still exist in the development process. If AI can reduce the risk involved in software development and take out some of the aspects of human error that is inevitable in project planning and cost estimation, software projects could have much greater chances of success. Many software engineers agree that AI could be invaluable in a support role in software development, even if they might not agree on how it could be implemented more fully into the entire process.

\bibliographystyle{ieeetr}
\bibliography{mybibliography}

\end{document}